\documentstyle[sprocl,epsfig]{article}

\def\Journal#1#2#3#4{{#1} {\bf #2}, #3 (#4)}

\def\be{\begin{equation}}
\def\ee{\end{equation}}
\def\bea{\begin{eqnarray}}
\def\eea{\end{eqnarray}}

\newcommand{\sax}{{\emph{Beppo}SAX} }
\newcommand{\nh}{cm$^{-2}$}
\newcommand{\lum}{erg~s$^{-1}$}
\newcommand{\kev}{\,\mbox{\scriptsize keV}}
\newcommand{\mum}{\:\mu\mbox{\scriptsize m}}

\begin{document}

\title{UNVEILING THE HIDDEN AGN IN THE MERGING STARBURST SYSTEM ARP 299 (IC~694+NGC~3690)}

\author{L. Ballo}

\address{Osservatorio Astronomico di Brera, via Brera 28\\20121 Milan, Italy\\E-mail: luballo@brera.mi.astro.it} 


\maketitle\abstracts{As part of a 
research of elusive AGN in a well defined sample of infrared 
selected galaxies (using {\em Beppo}SAX, Chandra,
{\em XMM-Newton} and INTEGRAL data) we have observed with \sax the 
merging starburst system Arp~299.
The broad-band (0.1--40 keV) coverage of this
observation 
clearly reveals, for the first time in this system,
the presence of a deeply buried AGN having an intrinsic luminosity of
 $L_{0.5-100\:\kev} \simeq 1.9 \times 10^{43}\:$\lum .}

\section{Introduction}
Studies of active objects at IR and X-ray wavelengths 
indicate that star-formation and AGN activity may be related (Fadda et al. 2002).
The trigger mechanism for both phenomena could be the interaction or the
merging of gas-rich galaxies.
This generates fast compression of the
available gas in the inner galactic regions, 
causing both the onset of a major starburst and the fueling of
a central black hole raising the AGN activity.
 
However the concomitant AGN and starburst activity is 
expected to happen in a high-density medium ($N_H \geq 10^{23-24}\:$\nh),
characterized by high dust extinction of the UV-optical flux and strong
photoelectric absorption of the soft X-rays (e.g. Fabian et al. 1998). 
Thus the study of these active phases in galaxies becomes very difficult; 
optical and even mid-/far-IR
spectroscopy may not be sufficient to disentangle starburst activity from 
AGN activity, which is actually best probed in the hard 
($E > 6\:$keV, 
in order to sample also the Fe K$\alpha$ line) X-ray energy 
band.

To search for hidden AGNs and to shed light on the starburst-AGN connection and its occurrence we have 
started
a systematic and objective investigation in hard ($E >6\:$keV) X-rays of
{\it a flux-limited sample of IRAS galaxies}.
The sample consists
of 28 galaxies selected from the {\it IRAS Cataloged Galaxies and 
Quasars} (see http://irsa.ipac.caltech.edu/)
as having $f_{60\mum} > 50\:$Jy
or $f_{25\mum} > 10\:$Jy. We 
stress here that no other selection criteria 
(e.g. established presence of an AGN, luminosities, IR colours, etc.) 
have been applied to the sample definition. 

\section{The merging starburst system Arp~299}

\begin{figure}[b!]
\begin{tabular}{cc}
\epsfig{figure=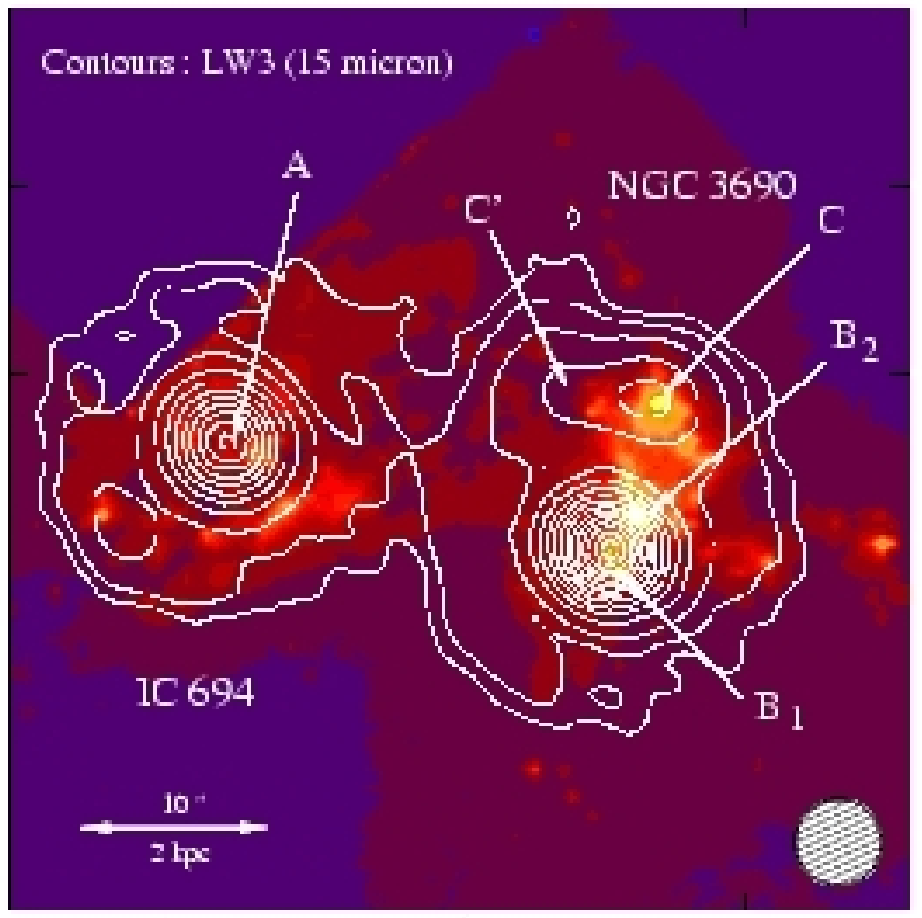, width=5cm}
& \epsfig{figure=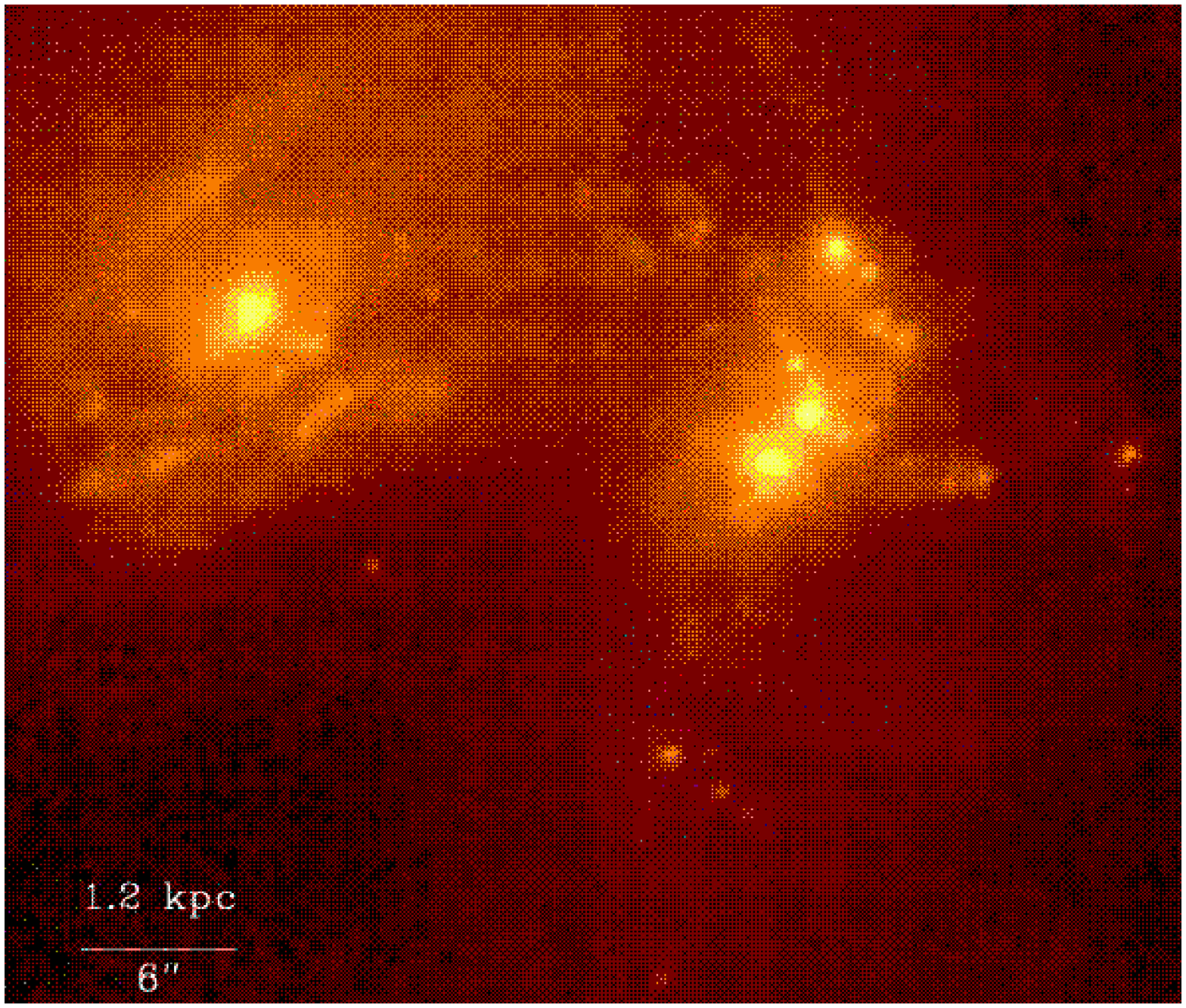, width=5.7cm}
\end{tabular}
\caption{The IR view of Arp~299. {\bf Left:} Mid-IR contours from ISO superimposed to the HST WFPC2 red image (adapted from Gallais et~al.~1999). {\bf Right:} HST NIC3 image of Arp~299 (adapted from Alonso-Herrero~et~al.~2000).\label{fig:system}}
\end{figure}

As a part of this research of elusive AGN, we have recently 
performed an X-ray observation with \sax of the merging starburst system Arp~299 
(= IC~694 + NGC~3690) located at D$\;=44\:$Mpc 
(z$\;=0.011$; Heckman~et~al.~1999).
Arp~299, a powerful far-IR system ($L_{43-123 \mum}=2.86 \times 10^{11}\:L_{\odot}$, dominating the bolometric luminosity), consists of two merging galaxies (NGC~3690 to the west and IC~694 to the east, see Figure~\ref{fig:system}), plus a compact galaxy lying to the northwest (Hibbard~\&~Yun~1999).
The centers of the two merging galaxies are separated by only $\sim 22^{\prime\prime}$.
Following the nomenclature of Winn-Williams~et~al.~(1991), the nucleus of the eastern galaxy (IC~694) is denoted~A, while NGC~3690 is resolved into a complex of sources: B1 and B2 to the south, and C and C$^{\prime}$ located $\sim 7^{\prime\prime}$ to the north of B2.

Both galaxies have been spectroscopically classified 
as starbursting from optical (Coziol~et~al.~1998) and mid-/far-IR 
(Laurent~et~al.~2000) data; previous observations in the X-ray band seems to support
these results (Zezas, Georgantopoulos and~Ward 1998, Heckman~et~al.~1999).

\begin{figure}
\vskip -0.7 true cm
\centerline{\epsfig{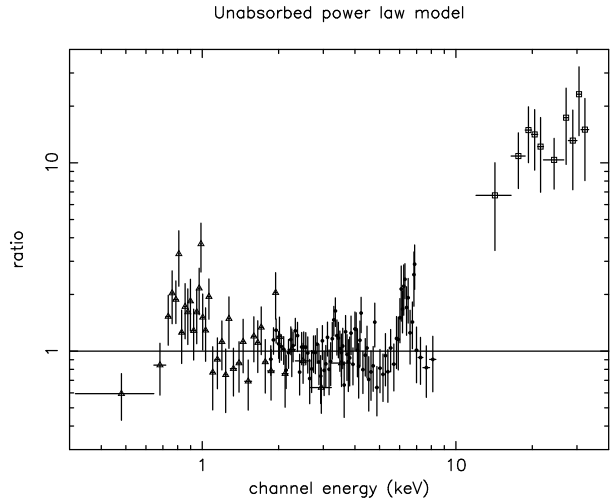}}
\vskip -0.3 true cm
\caption{Ratio between the unabsorbed power-law model and the 
\sax LECS (open triangles),
MECS (filled circles) and PDS (open squares) data.
\label{fig:unabspl}
}
\end{figure}

The \sax observations unveil however for the first time the presence of a
strongly absorbed AGN in this system 
(hints of the 
presence of an obscured AGN in the B1 nucleus of 
Arp~299 have been also recently inferred from near-IR 
spectropolarimetry; see R.~Maiolino~2002).
In Figure~\ref{fig:unabspl} we report the ratio between \sax data and the best-fit
 unabsorbed power-law model.
The residuals at $\sim 0.8\:$keV suggest the  presence of a soft thermal component,
which is a characteristic signature in all known
starburst galaxy X-ray spectra.
A line-like feature at $\sim 6.4\:$keV and a big bump in the PDS energy range
(10--40~keV) are clearly evident.
The simultaneous occurrence of these two characteristics is the
distinctive spectral signature of a heavily obscured AGN.

The spectral analysis (reported in Della~Ceca~et~al.~2002) reveals a 
deeply buried ``Compton-thick AGN'', with a column density of 
$N_H = 2.5\times 10^{24}\:${\nh}
and an intrinsic (i.e. unabsorbed) luminosity of
$L_{0.5-100\:\kev} \simeq 1.9\times10^{43}\:$\lum.

\begin{figure}
\vskip -0.7 true cm
\centerline{\epsfig{figure=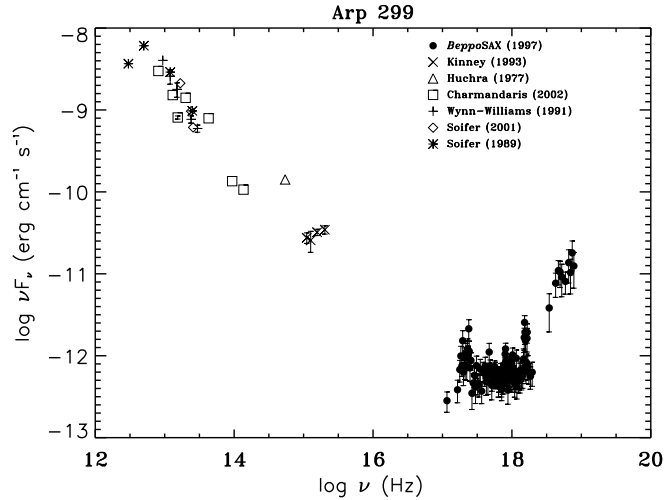, angle=90, width=9cm}}
\vskip -0.4 true cm
\caption{Broad-band spectral energy distribution of Arp~299. \label{fig:sed}
}
\end{figure}

The intrinsic X-ray luminosity of the AGN is a factor $\sim$ 50 less than 
$L_{FIR}$ ($\sim 10^{45}\:$\lum; the X-ray/optical/IR spectral energy 
distribution of Arp~299 is reported in Figure~\ref{fig:sed}).
Therefore the total FIR luminosity of the system cannot be 
entirely associated to the AGN, 
even assuming an AGN UV~luminosity 
a factor $\sim\:$10 greater than the X-ray luminosity (as
observed in QSOs, e.g. Elvis~et~al.~1994).
This suggests that the bulk of the FIR emission  
is due to the starburst, in agreement with the results
obtained by Laurent et al. (2000) using 5-16$\,\mu$m ISOCAM data.

The spatial resolution of the \sax instruments prevents us from 
localizing this AGN. To this purpose, X-ray observations from the Chandra and
{\em XMM-Newton} satellite have been retrieved from the public archive. 
From a preliminary analysis of the {\em XMM-Newton} data a strong line at
$\sim 6.4$ keV is clearly 
present in NGC~3690 and maybe also in IC~694, suggesting the possible presence 
of two AGNs.

\newpage

This work is part of a collaboration with R. Della Ceca, F. Tavecchio, L. Maraschi, 
P. O. Petrucci, L. Bassani, M. Cappi, M. Dadina, A. Franceschini, G. Malaguti, G. G. C. Palumbo and M. Persic.
Partial finantial support from ASI (I/R/073/01) and MIUR (Cofin-00-02-004) is acknowledged.

\end{document}